\documentclass[sigconf,nonacm]{acmart}

\makeatletter
\def\@ACM@bibliography@startnumber{1}
\makeatother
\title{Echoes of the Land: An Interactive Installation Based on Physical Model of Earthquake}

\author{Ivan C. H. Liu}
\orcid{0000-0002-9226-9765}
\affiliation{
  \institution{Future Narratives Lab\\Institute of Applied Arts\\National Yang Ming Chiao Tung University,}
  \city{Hsinchu}
  \country{Taiwan (R.O.C)}
}
\email{ivanliu@nycu.edu.tw}
\author{Chung-En Hao}
\affiliation{
  \institution{Future Narratives Lab\\Institute of Applied Arts\\National Yang Ming Chiao Tung University,}
  \city{Hsinchu}
  \country{Taiwan (R.O.C)}
}
\author{Jing Xie}
\affiliation{
  \institution{Future Narratives Lab\\Institute of Applied Arts\\National Yang Ming Chiao Tung University,}
  \city{Hsinchu}
  \country{Taiwan (R.O.C)}
}

\begin{document}

\begin{abstract}
\emph{Echoes of the Land} is an interactive installation that transforms seismic dynamics into a multisensory experience through a scientifically grounded spring-block model. Simulating earthquake recurrence and self-organized criticality, the work generates real-time sound and light via motion capture and concatenative granular synthesis. Each block acts as an agent, producing emergent audiovisual cascades that visualize the physics of rupture and threshold behavior. This work exemplifies the amalgamation of scientific knowledge and artistic practice, opening new avenues for novel forms of musical instrument and narrative medium, while inviting further investigation into the intersection of emergent complexity, aesthetics and interactivity.
\end{abstract} 

\maketitle

\section{Introduction}
Humans have long been drawn to the expressive vitality of natural phenomena. Clouds drifting, waves crashing, trees swaying---landscapes in motion do more than simply represent the environment. They evoke emotional resonance and perceptual depth, cultivating states of awe, curiosity, and contemplation. Recent findings in neuroaesthetics suggest that humans exhibit stronger aesthetic responses to dynamic natural scenes compared to static ones, underscoring the deep cognitive and emotional attunement we have to nature’s temporal unfolding \cite{zhao2020neural}. This inherent responsiveness to change and complexity forms a fertile ground for artistic inquiry.

From flocking of birds and ant colonies to traffic flow and financial markets, complexity is prevalent in both nature and society. Artists and scientists alike have been captivated by the emergent patterns arising from such systems, where patterns that are neither wholly random nor strictly ordered, but instead arise from distributed, rule-based interactions across scales. As early as the mid-20th century, researchers began investigating the aesthetic dimensions of complexity. Taylor and colleagues, for instance, revealed the fractal nature of Jackson Pollock’s poured paintings, linking their visual appeal to specific fractal dimensions \cite{taylor2011perceptual}. Such studies illuminate how complexity is not merely a scientific construct, but an aesthetic one, shaping how we perceive, process, and find meaning in visual experience. Today, with the advent of computational tools and algorithmic systems, artists are increasingly equipped to simulate natural processes and generate complex forms, engaging with concepts such as emergence, self-organization, and criticality \cite{mccormack2001art}. These developments invite a rethinking of the boundaries between art and science, not as opposing domains, but as overlapping inquiries into structure, transformation, and perception. Within this context, the simulation of dynamic natural systems becomes not just a technical exercise, but a generative aesthetic strategy, one that brings the intangible logic of the world into embodied, multisensory form.

A complex system comprises an ensemble of interacting elements that result in various collective phenomena, including fractal geometry, self-organization, threshold behavior and nonlinear dynamics. These emergent phenomena are absent when the system is reduced to its individual elements. Scientists have studied and theorized the underlying mechanisms of complex systems, which can be used to simulate or predict complex events. By abstracting a physical phenomenon into simple models, one can generate new representations that maintain the same physical principles but exhibit entirely different appearances. The present work utilizes the spring-block model from earthquake phenomena to create artistic representations. The phenomenon is a manifestation of self-organized criticality (SOC), a fundamental principle in complex systems science, describing how certain systems evolve to a critical state where small perturbations can lead to large-scale cascading effects \cite{brown1991simplified}. To our knowledge, visualization of SOC has only been performed for scientific purposes, not in an artistic context. Indeed, in a recent review article, Koentiz et al. addressed the need for new narrative representations of complexity and elucidated the associated challenges \cite{koenitz2022interactive}. Our objective is to create a new approach for artistic representation for viewers to experience and to explore the aesthetics of emergence, and present interesting findings from the creative process.

\begin{figure}
    \centering
    \includegraphics[width=\linewidth]{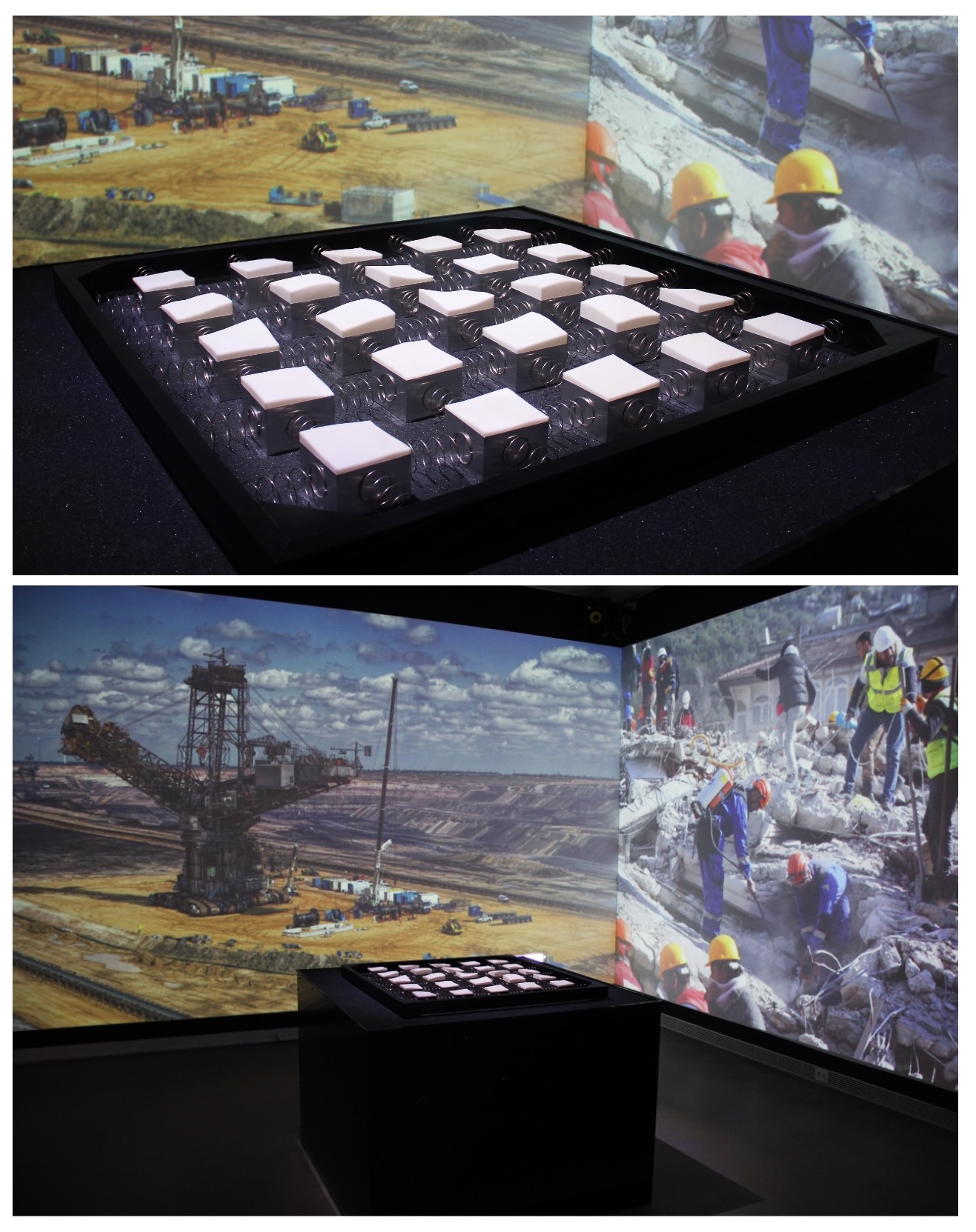}
    \caption{Upper: Close-up image of \emph{Echoes of the Land} showing the spring-block system; Lower: Spatial setup of the artwork including the immersive projection walls. (\copyright Ivan Liu)}
    \label{fig:1}
\end{figure}

This article elucidates the fundamental physics underlying the spring-block model and its application in creating the artwork \emph{Echoes of the Land} (\emph{EOL}) shown in Fig. \ref{fig:1}. We detail the technical aspects of the installation, including system design, interactive sound synthesis, light interaction and interactive visuals. The integration of complex physics with new media technologies gives rise to compelling aesthetic experiences. Moreover, it can be used to narrate global emerging issues such as anthropogenic seismicity \cite{liu2024narrating}. We also discuss how the present work relates to previous works on agent-based artworks as well as studies on aesthetics of complexity. This work introduces a novel approach to creating tangible interactive artworks based on complex physics.

\section{Self-Organized Criticality in Art}
Self-organized criticality is a class of complex phenomena known for its intriguing characteristics, exemplified by events such as avalanches and earthquakes. Conceptualized originally by Per Bak, Chao Tang and Kurt Wiesenfeld, their theory has been applied across many scientific disciplines \cite{bak1987self}. The visualization from SOC phenomena has mostly been for scientific purposes. However, recent artistic endeavors by Ivan Liu have employed materials such as rice grains, crop seeds and waste pills have been employed to create art installation \emph{The Rice-Pile Model}, based on avalanche mechanisms \cite{liu2024narrating, liu2023complexity}. The work demonstrated that the cascading dynamics of particles can generate captivating visual and auditory effects, presenting significant potential for artistic expression.

Earthquakes are self-organizing critical events resulting from the interactions of tectonic plates. Stress accumulated in the environment exerts pressure on the constituent terrain blocks, and when it surpasses a threshold, the stress redistributes through a series of tectonic movements that propagate throughout the system. These events release bursts of energy accumulated over time. The spring-block model simplifies the physical mechanism of earthquakes, serving as a theoretical framework to simulate their recurrence and capturing many of their physical properties \cite{brown1991simplified}.

\section{Distributed Agencies and Emergent Ecologies: From \emph{Sympathetic Sentience} to \emph{Echoes of the Land}}

The rise of agent-based modeling as an artistic methodology has opened new paradigms for exploring emergence, self-organization and systemic behavior in media art. By structuring activity through decentralized rule-based interactions, such works foreground the aesthetics of complexity and the dynamic properties of autonomous systems. This section examines a lineage of artworks that utilize these ideas and positions \emph{EOL} within this tradition, emphasizing its foundation in a scientifically validated geophysical model and its unique engagement with self-organized criticality

Simon Penny’s \emph{Sympathetic Sentience} (1993--1996) exemplifies early investigations into emergent sonic ecosystems driven by autonomous agents. Each unit in the system emits and receives rhythmic signals via infrared communication, forming a distributed acoustic network that reorganizes itself in response to human presence. The behavior of the system is not scripted but evolves dynamically through feedback, highlighting key features of complex adaptive systems such as nonlinearity, emergence and resilience. So Kanno’s \emph{Lasermice} (2019) advances this logic in a robotic context. In this work, small mobile robots synchronize their laser emissions through local sensing and signaling, forming an ephemeral, light-based communication web. As in \emph{Sympathetic Sentience}, external disturbances, especially from viewers, interrupt and reshape the system’s evolving structure, reinforcing the role of the audience as an ecological agent within the work. Ken Rinaldo’s \emph{Autopoiesis} (2000), Choeur \emph{Synthétique} (2023) by Muhanad Alkilabi and co-creators, and Tim Blackwell’s \emph{Swarm Music} (2002) continue this trajectory by deploying networked agents to generate complex sonic or kinetic environments. These works prioritize behavior over form and structure over objecthood, making visible and audible the real-time unfolding of multi-agent systems. While they successfully illustrate collective behavior and emergent complexity, they do not explicitly engage with the statistical or dynamic regimes associated with critical phenomena.

Building upon these precedents, \emph{EOL} introduces a unique intervention by grounding its emergent system in a rigorous scientific model used in seismology to simulate earthquake recurrence. The system exhibits stick-slip dynamics and spontaneous, large-scale events (earthquakes) arising from slow, local interactions, leading to the characteristic of SOC. 

In \emph{EOL}, each agent corresponds to a block within this model and operates according to physical parameters such as coupling stiffness, friction and external loading rate. As the system evolves, slow stress accumulation leads to sudden ruptures and propagating chains of motion, generating audiovisual output in response to simulated seismic activity. These cascades, ranging from minor tremors to large-scale ruptures, are not pre-programmed but emerge from the system’s internal nonlinear dynamics, forming an artistic embodiment of criticality absent in the other works discussed.

Moreover, while previous agent-based artworks employ swarm logic or localized rule systems to generate interaction, \emph{EOL} explicitly engages with Langton’s “edge-of-chaos” regime, where order and disorder coexist, life and creativity thrive, and aesthetics emerges \cite{langton1990computation, goodwin2001leopard}. This positions the work within a theoretical space shared by complex systems science, far-from-equilibrium thermodynamics and critical state theory. The inclusion of SOC dynamics allows the work not only to visualize complexity, but to materialize the physics of emergence, offering a sensory encounter with the same principles that underlie natural systems such as forest fires, neuronal avalanches and earthquakes.

This integration of agent-based modeling, rigorous scientific simulation and aesthetic expression makes \emph{EOL} a distinct contribution to the discourse on complexity in media art. It shifts the role of the artist from designer of interactive behaviors to curator of phase transitions, inviting audiences into a system that behaves with the volatility, memory and scale-invariance of critical systems in nature. In doing so, the work expands the conceptual vocabulary of agent-based art, pointing toward new possibilities for embedding scientifically grounded complexity into affective, embodied and environmentally resonant media experiences.

\section{Earthquake Physics}

\begin{figure}
    \centering
    \includegraphics[width=\linewidth]{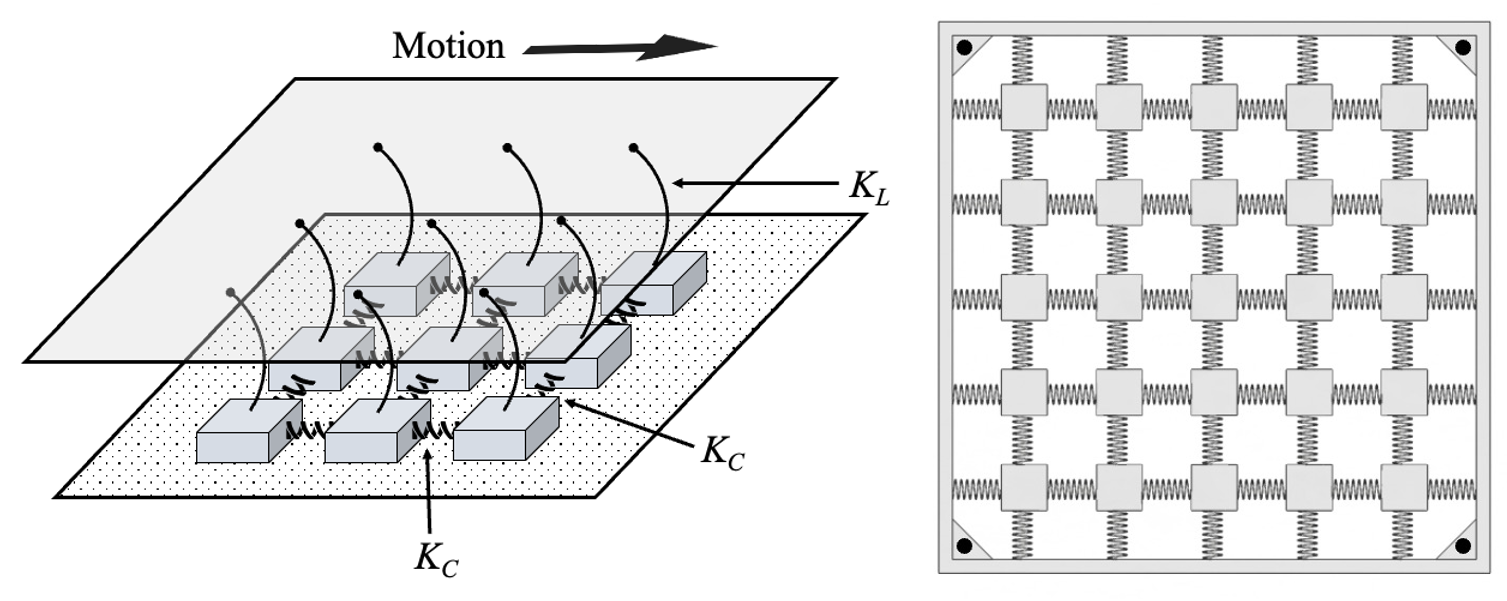}
    \caption{Left: The two-dimensional spring-block model for simulating the earthquake recurrence proposed by Brown et al. \cite{brown1991simplified} Right: The illustration of the adapted 5×5 spring-block system. The dots at the corners indicate the locations of caster wheels. The enclosing frame provides an interface for interaction with the earthquake system.}
    \label{fig:2}
\end{figure}

The spring-block model for simulating the recurrence of earthquakes was initially proposed by Burridge and Knopoff in one dimension and later extended to two dimensions by Brown et al. \cite{burridge1967model, brown1991simplified}. This model features two-dimensional arrays of blocks interconnected by springs with stiffness $K_C$ on a frictional surface. Each block is also connected to an overlying suspended plate by springs with stiffness $K_L$, which move horizontally, as illustrated in Fig. \ref{fig:2}. Typically, $K_L$ is much smaller than $K_C$, simulating a weaker shearing force from the overlying plate compared to the interaction between adjacent blocks. 

The recurrence of earthquakes is initiated by moving the top plate in a horizontal direction. The springs, $K_L$, pull the blocks along with the plate. However, due to static friction between the blocks and the surface, the blocks remain stationary until the pulling force overcomes the static friction. When the external force exceeds this threshold, at least one block begins to slip. The slippage then triggers an onset of sequential movements propagating throughout the network of blocks via the springs $K_C$. The motion ceases once the stored potential energy falls below the threshold, until it is reinitiated by the moving plate.

The total force acting on a block at the i-th row and j-th column is \cite{brown1991simplified}

$F_(i,j)=K_L l_(i,j)+K_C [4l_(i,j)-l_(i-1,j)-l_(i+1,j)-l_(i,j-1)-l_(i,j+1) ]$

where $l_(i,j)$ is the displacement of the block at $(i,j)$ from the previous moment with a fixed timestep.
Worldwide monitoring of seismic activity has led to extensive data collection on earthquake occurrences. Research has demonstrated that the number of earthquakes per unit time interval, $N(A)$, with a rupture area greater than or equal to A follows a power-law distribution \cite{hergarten2003landslides},$N(A)~A^(-b)$, where $b$ is a parameter typically between 0.8 and 1.2 depending on the region. This equation reveals that all earthquake sizes adhere to the same power-law relationship, indicating scale invariance, which is a key characteristic of self-organized criticality.

\section{Mechanical Design}

\begin{figure*}
    \centering
    \includegraphics[width=\textwidth]{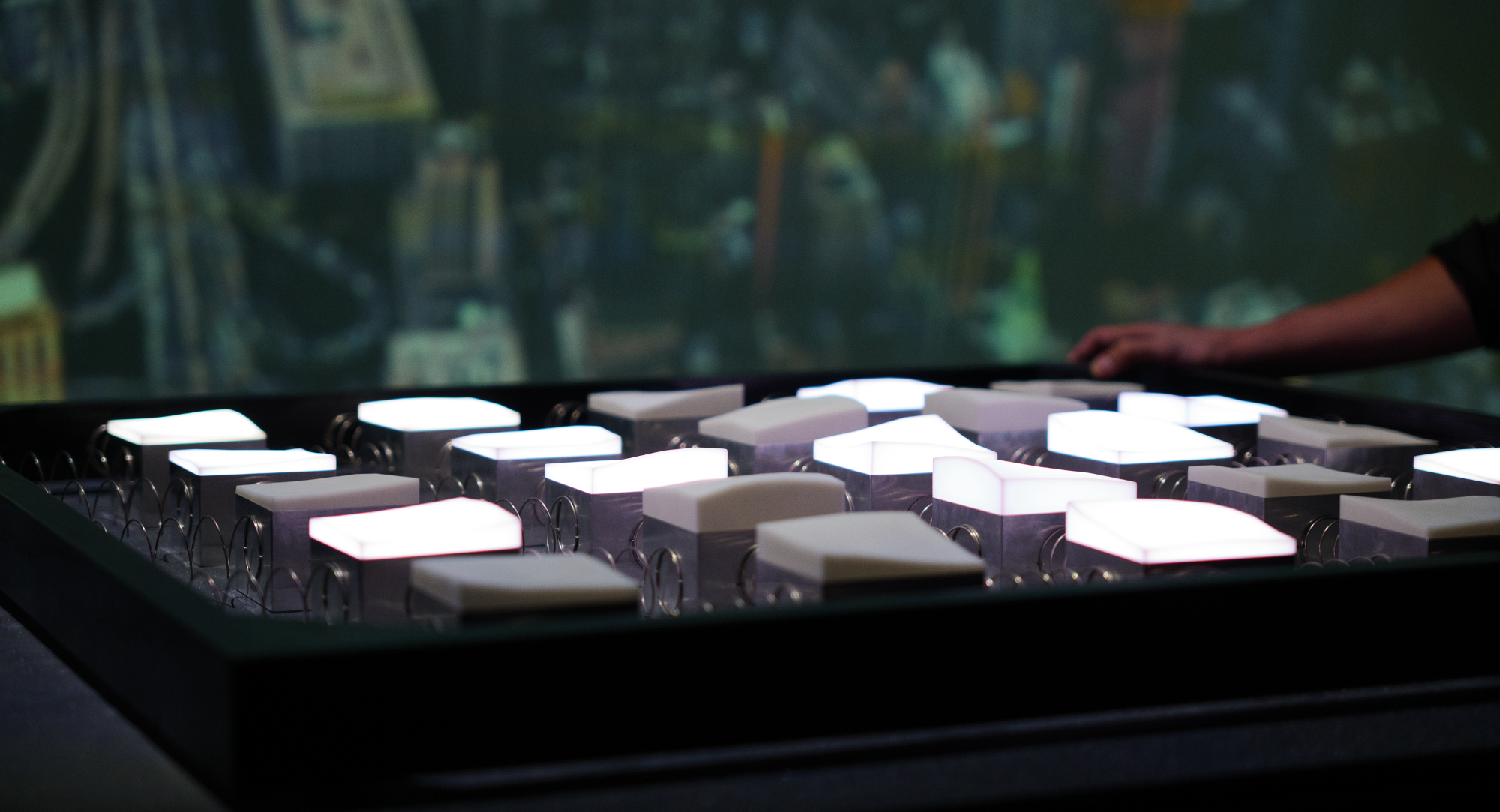}
    \caption{Shows an audience interacting with the installation by moving the frame of the spring-block ensemble, generating light cascades. (\copyright Ivan Liu)}
    \label{fig:3}
\end{figure*}

To generate the stick-slip movement using the spring-block ensemble while maintaining clear visibility, the overlying plate is modified into an enclosing frame containing 25 blocks arranged in a $5\times5$ array, as illustrated in Fig. \ref{fig:2}. This adjustment removes the visual obstruction on top, allowing the audience to directly observe the movement of the blocks. However, this simplification isolates the nine central blocks from the frame, causing a latency in their response to the moving frame. Despite this, the dynamics of the modified setup are sufficiently similar to the original configuration to produce an engaging visual and audio experience. The deviation from the original spring-block model increases with the size of the block ensemble due to the greater number of central blocks that are not coupled to the frame.

The four corners of the enclosing frame are equipped with caster roller wheels, allowing the frame to move smoothly in all directions without adding extra friction to the system. This design also provides the participants with greater freedom to express themselves, as they can push, pull, or rotate the ensemble (see Fig. \ref{fig:3}). Additionally, multiple participants can engage with the installation simultaneously, as it is identical on all four sides.

The dimensions of the square frame, the blocks and the springs are 960 cm, 8.4 cm and 9 cm, respectively. Each block weighs approximately 250 g. The spring-block system is placed on a square platform measuring 140×140 cm, covered with 120-grit sandpaper. The bottom sides of the blocks are also covered with sandpapers of the same grade. After experimenting with various surfaces, including wood, baize and other fabrics, we found that this setup creates the most appropriate friction dynamics to accurately simulate earthquake recurrence.

\section{System Design}

\begin{figure}
    \centering
    \includegraphics[width=\linewidth]{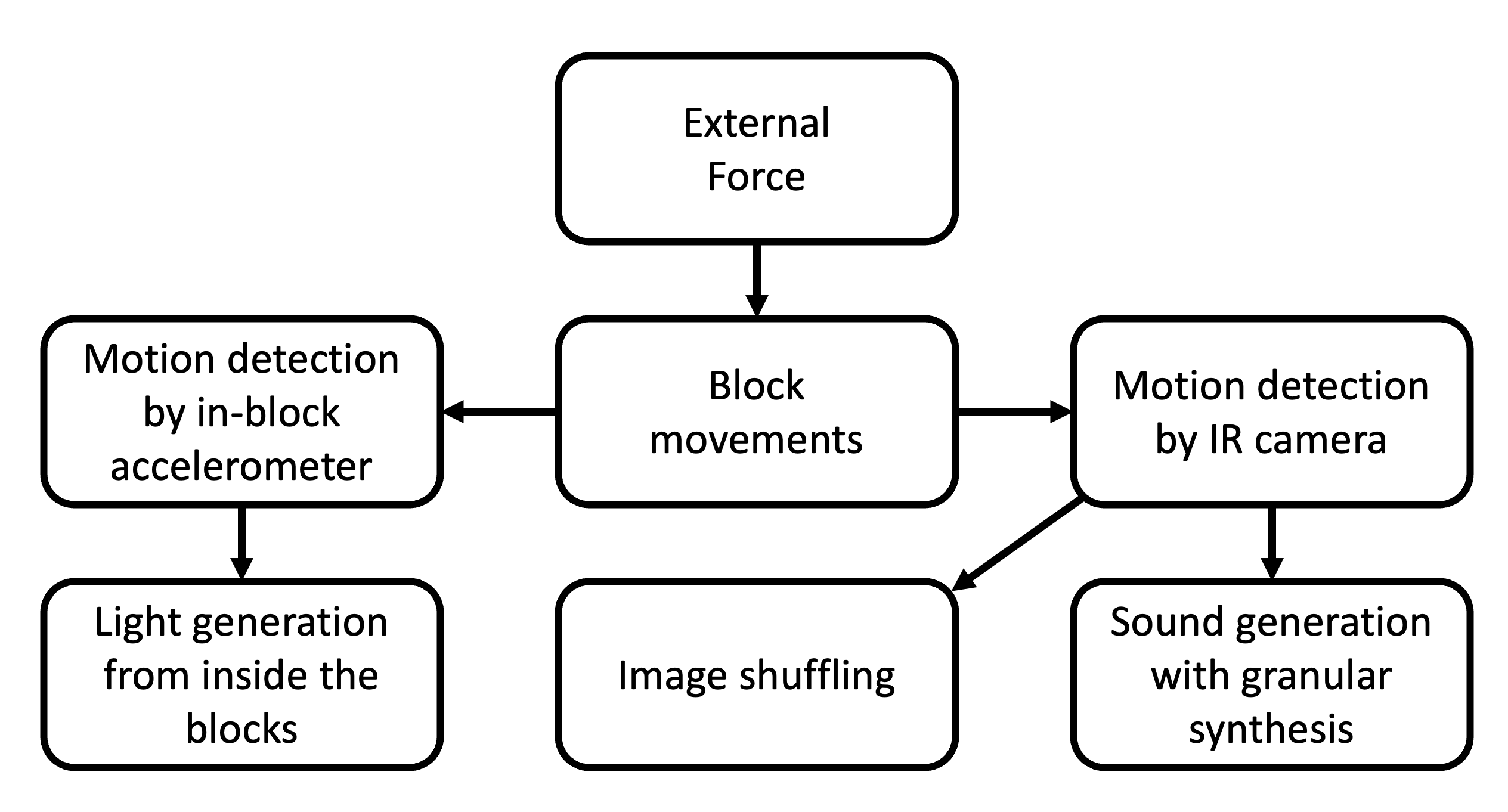}
    \caption{A flowchart showing how the system respond to the human interaction.}
    \label{fig:4}
\end{figure}

\begin{figure}
    \centering
    \includegraphics[width=\linewidth]{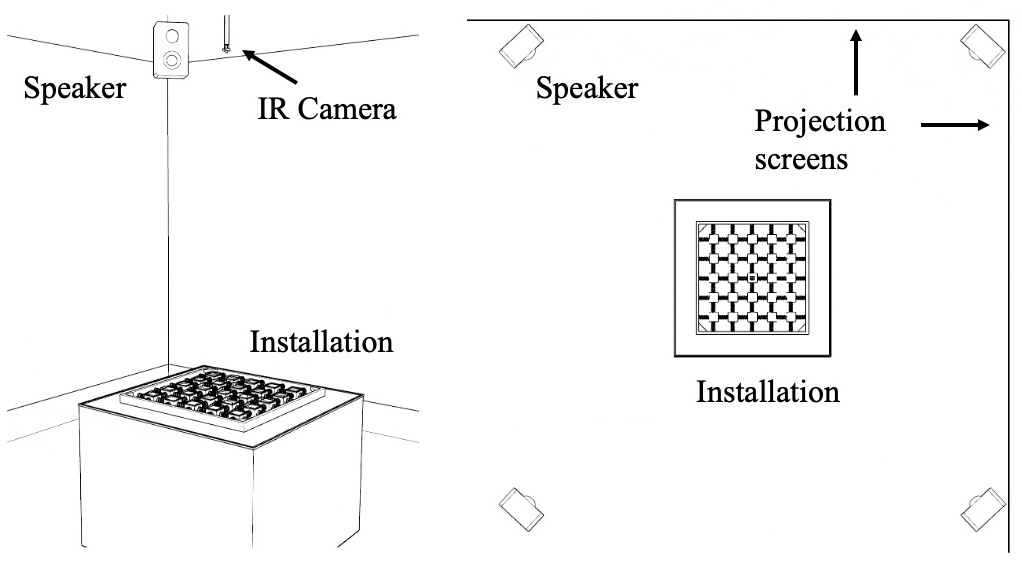}
    \caption{Schematic diagrams of the spatial setup in different views from 45 degree (left) and top view (right).}
    \label{fig:5}
\end{figure}

Figure \ref{fig:4} illustrates the system workflow of the installation. The audience interacts with the installation by moving the enclosing frame horizontally through pushing, pulling, or rotating as desired. This movement exerts force on the spring-block ensemble, initiating a stick-and-slip motion.
To detect the motion of the blocks, infrared (IR) LEDs are installed inside each block. An infrared camera, suspended above the table (Fig. \ref{fig:5}), captures the movements. The top-left image in Fig. \ref{fig:6} (upper) shows the live feed of a person interacting with the spring-block system. Using an IR detection system minimizes interference from ambient or general exhibition lighting. As seen in the photo from the actual exhibition (Fig. \ref{fig:1}) and the IR live image (Fig. \ref{fig:6}), even with environmental lighting, IR signals from the blocks, appearing as bright purple squares, are consistently strong, ensuring reliable and stable motion tracking. The computer records the coordinates of the blocks and analyzes their motion using a frame-difference algorithm in a MAX/MSP program. The motion data is sent to a CataRT module for real-time concatenative granular synthesis to generate sound \cite{schwarz2006real}. Simultaneously, the motion data triggers a sequence of rapid image montages, which are explained in the following sections.

\section{Real-Time Sound Generation with Concatenative Granular Synthesis}

To generate a natural and realistic earthquake sound effect in real time, we employ concatenative granular synthesis, a technique developed by D. Schwarz et al \cite{schwarz2006real}. This technique enables real-time spatial triggering of selected sound grains based on their characteristics. It performs a concatenative analysis of segmented sounds, arranging the segments into a two-dimensional descriptor space according to properties such as pitch, volume and timbre. By mapping movements in real space to this descriptor space, sound segments are triggered continuously, producing a natural and immersive auditory experience.

\begin{figure}
    \centering
    \includegraphics[width=\linewidth]{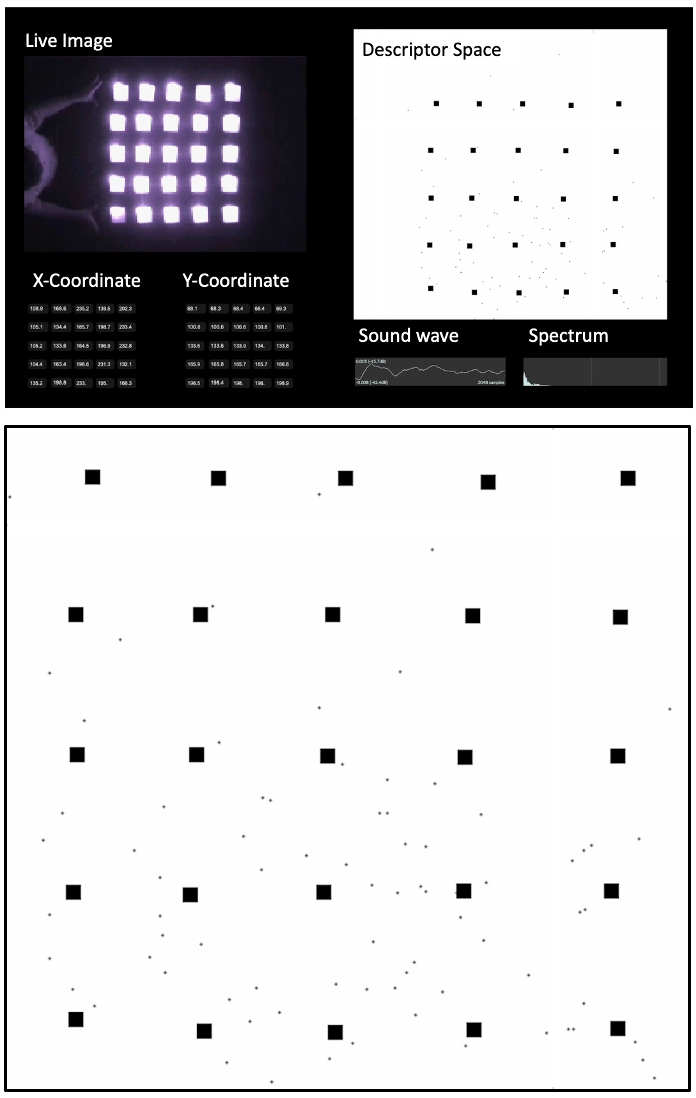}
    \caption{Upper: The panel displayed on a monitor showing the live image from the IR camera, the x and y coordinates of the blocks, the descriptor space, the live audio waveform and spectrum. Lower: The descriptor space showing the sound grains (scattered dots) of the input sound sample. The squares are the positions of the blocks mapped from real space. (\copyright Ivan Liu)}
    \label{fig:6}
\end{figure}

The sound sample used is that of an iceberg breaking, as it contains an appropriate balance of low- and mid-frequency components, resulting in a synthesized sound that gives an auditory impression of the earth moving and trembling. Figure \ref{fig:6} (lower) shows the descriptor space created from granular synthesis, with the centroid mean and periodicity mean as the x- and y-axes, respectively. These axes are chosen to evenly distribute the sound grains within the interactive region, allowing any block's movement to trigger sound. Alternatively, the grains can be repositioned evenly within the descriptor space by manually editing their coordinates. The black squares in Fig. \ref{fig:6} (lower) represent the mapped positions of the blocks in real space. Each block triggers a sound when it moves into proximity with a sound grain, with an adjustable range.

\section{Visual}

\begin{figure}
    \centering
    \includegraphics[width=\linewidth]{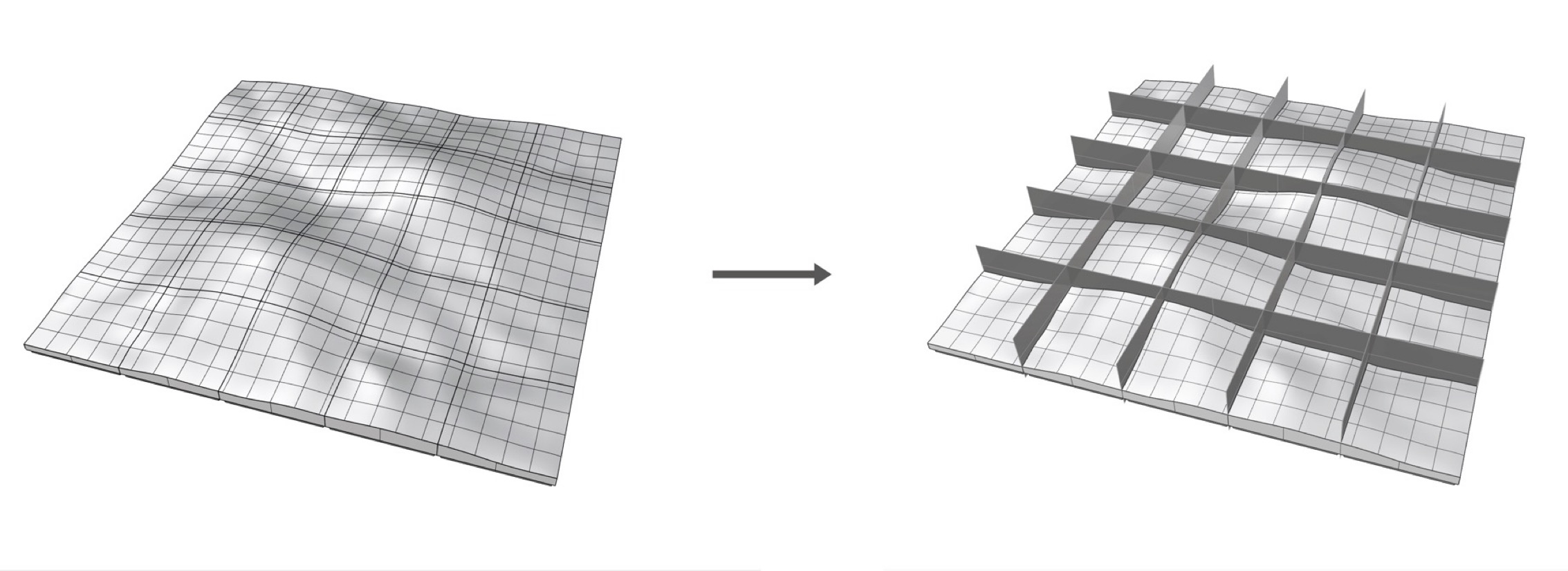}
    \caption{ The semi-translucent covers of the blocks are segments from a continuous virtual 3D terrain.}
    \label{fig:7}
\end{figure}

\begin{figure}
    \centering
    \includegraphics[width=\linewidth]{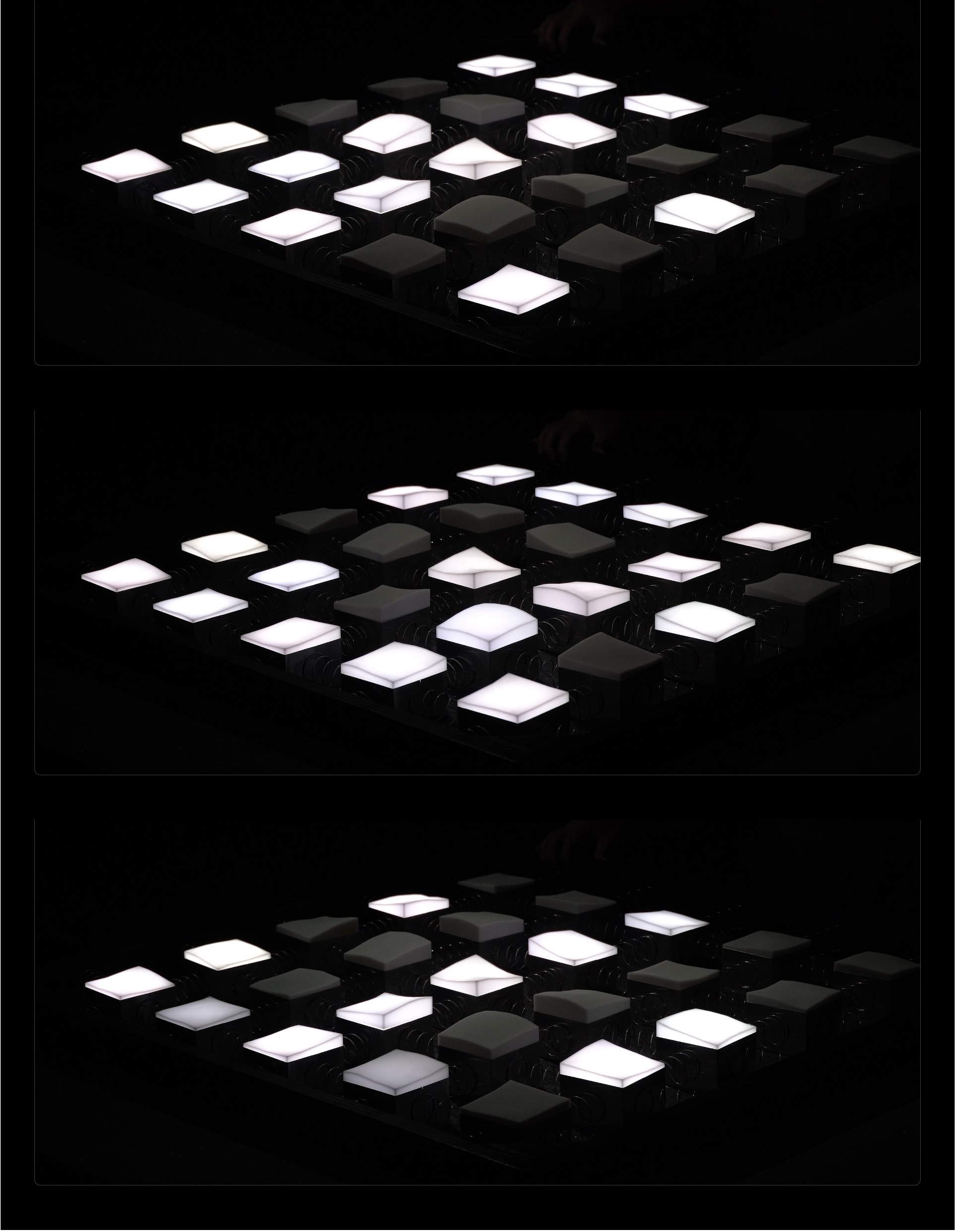}
    \caption{Light sequence of three image frames roughly 0.5 seconds apart from top to bottom. The ensemble was pushed from the bottom-left to the top-right. (© Ivan Liu)}
    \label{fig:8}
\end{figure}

Two interactive visual elements are activated when the blocks move: light generation from the blocks and the rapid shuffling of the image montage, as illustrated in Fig. \ref{fig:4}.
Each block emits a flash of white light through a semi-translucent cover when its movement is detected by an internal accelerometer. These covers are 3D-printed segments from a continuous virtual terrain (shown in Fig. \ref{fig:7}), creating the impression that participants are shifting the landscape, enhancing the contextual perception of the artwork. The light flash not only indicates movement but also generates a cascading lighting effect triggered by nearest-neighbor interactions of the spring-block ensemble, as shown in Fig. \ref{fig:8}. While the initial pattern may appear random, the emerging dynamic light patterns soon resemble natural phenomena such as lightning or the spread of a wildfire.

In order to mediate between the viewers’ perception and the contextual meaning of anthropogenic seismicity, we designed an L-shaped immersive projection space (Fig. \ref{fig:5}). One side of this space displays scenes of human industrial activities, such as hydroelectric dams, hydrofracking and oil recovery, while the other side shows images of potential consequences, such as the destruction of houses and roads. When block movements are detected by the IR camera, the paired images in the database shuffle rapidly at 0.5-second intervals, accompanied by the trembling sounds generated by granular synthesis. The images freeze when the blocks are still. 
This display method highlights the causality between anthropogenic activities that can induce seismicity and their potential disastrous events. The rapid shuffling of images creates a dreamlike ambience and tension that captivates viewers. The combination of immersive visual images and sound creates an aesthetic experience of sublimity that transcends verbal description.

\section{Complexity, Order and Aesthetics}

The representation of the spring-block ensemble as lit and dark blocks resembles a tow-dimensional cellular automata. Indeed, the physical features of spring-block model have been studied by scientists in the context of cellular automata \cite{olami1992self}. Since there have been extensive theoretical and experimental investigations into the aesthetics of cellular automata, it is valuable to apply their results to shed light on how audiences might experience certain elements of \emph{EOL} aesthetically. The discussion in this section also demonstrates the potential of the spring-block ensemble as a tangible interactive medium for investigating the aesthetics of cellular automata.

In 1933, Birkhoff famously defined aesthetic measure, $M=O/C$ , as the ratio between order, $O$ , and complexity, $C$ \cite{birkhoff1933aesthetic}. According to this relationship, a higher degree of order increases aesthetic value, while a higher degree of complexity decreases it, and vice versa. Although later experimental investigations by other researchers produced contradictory results regarding this simple mathematical relationship, Birkhoff’s methodology and analysis provided an invaluable framework for quantitatively investigating aesthetics for future researchers. 

From this perspective, an artwork can be ordered or disordered (random), simple or complex, and these elements often complement each other to enhance the viewer’s aesthetic experience. As Arnheim noted, “Complexity without order produces confusion. Order without complexity causes boredom” \cite{arnheim1966toward}. Complexity can be defined by aspects related to quantity and variety of information in an artwork, while order can be defined by aspects related to structure and organization of information \cite{van2020order}. A highly complex image may feature numerous turns, intersecting lines, or colors, whereas a highly ordered image may display repetition, similarity, symmetry, or sequence. 

As noted by Birkhoff, different types of stimuli might lead viewers to associate them differently based on their subjective perceptions \cite{birkhoff1933aesthetic}. Consequently, comparing studies involving similar types of stimuli yields more meaningful insights. The binary visual patterns generated by the block lighting in \emph{EOL} are similar to those studied by Chipman \cite{chipman1977complexity}, and Krpan and van Tilburg \cite{krpan2022aesthetic}, providing a framework for discussing the aesthetics of the generated patterns. The latter proposed the Aesthetic Quality Model (AQM), which posits that the artistic quality of an artwork is perceived based on the combination of its complexity and randomness. They conducted an experiment in which participants viewed black-and-white patterns arranged on a two-dimensional 6×6 grid with varying degrees of complexity and randomness and rated the patterns based on their perceived aesthetic value. The study found that patterns with low randomness and high complexity were rated as more aesthetically appealing \cite{krpan2022aesthetic}. This scenario inclines with the self-organizing patterns of \emph{EOL}, which are inherently complex, yet nonrandom, suggesting greater aesthetic appeal for the viewers.

\section{Fractal Music}

Through the sonification of block movements, the installation generates a time series of earthquake sounds with multiscale microstructures corresponding to the recurrence of the stick-slip movement, which exhibits a power-law distribution as previously mentioned. This power-law regularity is found in both human and non-human music, suggesting a natural human affinity for music with such structures \cite{VOSS1975, jermyn20231}. These sounds exhibit a power density proportional to $1/f$, where $f$ represents the sound frequency. Noise with such power-density spectrum is sometimes referred to as fractal noise. The installation thus offers a novel tangible interface for generating “fractal music”.

\section{Conclusion and Future Prospects}

In conclusion, we have developed an interactive installation that incorporates sound and light inspired by the physics of earthquake recurrence. This installation exemplifies the fusion of scientific modeling, artistic expression and narrative, transforming abstract earthquake dynamics into a tangible and immersive experience. It simultaneously generates scientific data and evokes artistic sublimity. The kinetic movement of an earthquake is simulated in real-time by a modified spring-block model, which triggers trembling sounds and collective light cascades through motion capture technologies. The resulting visual and audio representations emerge from the underlying complex dynamics. In this sense, the complex model acts as a generator for the visual and sound elements. 
The sonic dimension is produced using concatenative granular synthesis, with the spatial and mechanical coordinates of the spring-block ensemble mapped to a sound descriptor space. The result is a dynamic, naturalistic sonic texture that reflects the physical and temporal structure of seismic events. The interactive lighting system enhances this relationship, providing the audience with an immediate visual correlation between structural stress, rupture and sonic articulation creating a multisensory experience of emergent criticality.
The present system design, as illustrated in Fig. \ref{fig:4}, offers a versatile framework applicable to a wide range of art projects. The multimedia content such as sound and visual images can be replaced to facilitate different contexts, and the mechanical design and shape can be modified according to the desired artistic concept.

The spring-block model can be interpreted as an agent-based system, where each block acts as a semi-autonomous agent responding to local stresses. This positions \emph{EOL} within a lineage of agent-based and emergent media artworks which explore distributed agency and self-organization. However, \emph{EOL} introduces a distinctive advancement: the integration of a real scientific model capable of exhibiting behavior near SOC. This threshold-driven dynamic has rarely been realized in an artistic context. It allows the work to embody not only aesthetic complexity but also a deep structural resonance with the dynamics of natural systems.
The current system is modular and highly adaptable, supporting various media contexts through interchangeable sound and visual elements, and it has been exhibited internationally, including IRCAM (Paris), LEV Festival (Gijón), ISEA25 (Seoul) and the Kaohsiung Museum of Fine Arts (Taiwan). Its potential extends beyond installation: the real-time interaction between the spring-block ensemble and human participants offers a new form of performative, complexity-based musical instrument, enabling the generation of “fractal music” through embodied control and feedback.
Drawing on frameworks such as the Aesthetic Quality Model and the concept of the edge of chaos, the audience is likely to find the patterns generated by \emph{EOL} aesthetically pleasing due to their inherent balance of complexity and disorder. It is interesting to point out that there has been limited research on the aesthetic appreciation of dynamical complex patterns. As mentioned at the beginning of this article, neuroaesthetic studies with landscape imagery suggest that humans find dynamic stimuli more beautiful than static ones. Temporal evolution may add further dimensions to the aesthetics of complex patterns.
We hope the concepts in this work inspire further artistic explorations. Ultimately, this project contributes to the field of complexity and agent-based art by offering a tangible, scientifically grounded approach to exploring threshold behavior, emergence and nonlinear interaction. It invites audiences to not only witness but also participate in a real-time, complexity-informed aesthetic system, and it offers a novel means of engaging with global issues such as anthropogenic seismicity and environmental disruption \cite{liu2024narrating}. 

\section*{Acknowledgements}
We thank Diemo Schwarz for his helpful advice on the use of granular synthesis, and Jon McCormack for helpful comments on earlier versions of the manuscript. This project was funded by the National Science and Technology Council (Taiwan) with grant no. MOST-111-2420-HA49-005 and NSTC 112-2420-H-A49-002.

\bibliographystyle{abbrv}
\bibliography{refs2}

\end{document}